\documentclass[conference]{IEEEtran}
\usepackage{graphicx}

\usepackage{cite}

\ifCLASSINFOpdf
\else
\fi

\usepackage{todonotes}

\begin{document}
%
\title{At Ease with Your Warnings:\\ The Principles of the Salutogenesis Model Applied to Automatic Static Analysis }

\author{\IEEEauthorblockN{Jan-Peter Ostberg, Stefan Wagner}
\IEEEauthorblockA{University of Stuttgart\\
Institute of Software Technology\\
Stuttgart, Germany\\
Email: \{Jan-Peter.Ostberg,Stefan.Wagner\}@informatik.uni-stuttgart.de}
}


%


\maketitle

\begin{abstract}
The results of an automatic static analysis run can be overwhelming, especially for beginners. The overflow of information and the resulting need for many decisions is mentally tiring and can cause stress symptoms.
There are several models in health care which are designed to fight stress. One of these is the salutogenesis model created by Aaron Antonovsky.
In this paper, we will present an idea on how to transfer this model into a triage and recommendation model for static analysis tools and give an example of how this can be implemented in FindBugs, a static analysis tool for Java.
\end{abstract}


%
\IEEEpeerreviewmaketitle

\section{Introduction}
Modern tools for automatic static analysis can provide a huge amount of information to the user (e.g.\ shown by Ruthruff et al.~\cite{ruthruff2008predicting}). In most cases this information can be filtered and modified by various options provided by the tool. The vast amount of options is hard to handle for beginners and even discouraging. Moreover, cutting through this jungle of findings and options of the tool costs a lot of time and cognitive performance.

As Schwartz~\cite{schwartz2004tyranny} explains, more options also mean more possible regret about the options not taken. So, I am never sure if I have chosen the best option until I have explored all possibilities by trial and error. This is not feasible for a large set of options, especially if the decision on one option unfolds another layer of options to choose from. As we know that, we accept a selection at some point in time. To avoid cognitive dissonance, a theory formulated by Festinger~\cite{festinger1962theory}, we might adapt our desire to match our selected options instead the actual best fit to our needs.

\subsubsection{Problem Statement}
To enable a better interaction between the user and the tool, the initial hurdle needs to be lowered and the amount of information besieging the user needs to be reduced. But how can we achieve this without frustrating the user? It is crucial to leave some control with the user. Whitworth~\cite{whitworth2005polite} calls this ``being polite".  Otherwise, the tool might trigger the \textit{learned helplessness} effect described by Maier and Seligman~\cite{maier1976learned}: Learned helplessness is caused by uncontrollable or seemingly uncontrollable events which could cause organisms to remain passive, like the rabbit caught in the headlights of a car. If situations like this are encountered repeatedly, organisms stop acting even if responding to the situation would be effective. In our case, if the worst comes to the worst,  this would lead the user either to not thinking about the results presented or avoiding the tool at all.

\subsubsection{Objective}
Therefore, our goal is to create an interaction model which helps the users of static analysis tools to triage the findings effectively and without stress. Additionally, we aim at continuously increasing the knowledge on the issues triaged and, thereby, improving future triage runs.

\subsubsection{Contribution}
In this paper, we propose such an interaction model which reduces the information overload and stress indicators with the aim to create a positive emotional connection with the tool. For that, we transfer the \emph{salutogenesis model}, first introduced by Antonovsky~\cite{antonovsky1979health}, to the area of static analysis tools as a way to focus their finding. We describe how this model can be used on the results of a static analysis tool and give a concrete example for the implementation with FindBugs.\footnote{http://findbugs.sourceforge.net/}

\section{The Model}
The original salutogenesis model was introduced by Antonovsky in the 1970s.  We will lean on Lindstr{\"o}m and Erikson~\cite{lindstrom2005salutogenesis} in the following description of the original model.  In contrast to the prevailing approach of understanding what makes people ill (\emph{pathogenesis}), Antonovsky wanted to know what factors have an influence on people staying healthy. He does not define health as a binary state (healthy or ill) but as a continuum ranging from ill health (\emph{dis-ease}) to total health (\emph{ease}). The ``ease" is influenced by the person's ability to comprehend the situation he or she is in which is  described in more detail by the abilities to assess, understand and find meaning in the person's living situation. Antonovsky called the state of mind created by these three abilities ''the sense of coherence''. So, if a person cannot understand, assess or find meaning in his or her living conditions, he or she will have a low sense of coherence or, in other words, is stressed, and so will be more likely to get ill. He or she will be dis\textit{eased}.

Antonovsky did not talk about individual abilities but circumstances that influence his model. While it may not be possible to change a person's abilities, the circumstances of a situation may well be changed. So he speaks of \textbf{comprehensibility} in a situation where the ability to understand the situation is needed. Often adding information is the key. For example, explaining to a child why something is dangerous increases the comprehensibility for the child.
Antonovsky speaks of \textbf{manageability} when the assessment of a situation is needed. For example, if there is too much work for a given time, prioritization would increase manageability.
\textbf{Meaningfulness} is necessary when finding meaning in the situation is needed. For example, if people feel a lack of meaning in their working life and might be depressed (diseased), it might help them to plan what future goals they want to reach or why their work is important to increase meaningfulness. 

\section{Related Work}

\subsubsection{Work Influencing Manageability}
Increasing manageability means reducing the amount of  information we need to process and being able to better focus our limited resources. Heckman~\cite{Heckman:2007:ARA:1349332.1349339} proposed a model based on code locality and developer information and Ruthruff et al.~\cite{ruthruff2008predicting} use logic regression models to find actionable warnings. Both contribute to manageability.

Time management is another aspect of manageability. Also for that aspect, there is work like that by Weiss et al.~\cite{Weiss:2007:LTF:1268983.1269017} who use existing reports in an issue tracking system to find similar tasks and use their average duration as a prediction, or by Giger et al.~\cite{Giger:2010:PFT:1808920.1808933} who used decision tree analysis to predict the effort and time to fix a bug.

Additionally, this aspect is influenced by HCI research with contributions like the work of Shneiderman~\cite{shneiderman1992designing} or Nielson~\cite{nielsen1994usability}. They cover topics like the optimal UI design for reducing information overload.

\subsubsection{Work Influencing Meaningfulness}
Increasing meaningfulness means adding information that puts the object lacking meaningfulness into perspective. While dealing with the results of static analysis tools this can be reached to some degree by the explanations given by the tools themselves. Yet, these explanations can be more confusing than helpful. Johnson et al.~\cite{johnson2013don} found in their study that ``19 out of 20 participants felt that many static analysis tools do not present their results in a way that gives enough information for them
to assess what the problem is.''

Another way to add meaning is by using a detailed quality model like \emph{Quamoco}~\cite{wagner:ist15} which explicitly connects the results of static analysis to quality attributes.

\subsubsection{Work Influencing Comprehensibility}
To increase comprehensibility usually means adding context to the finding. This aspect of salutogenesis is mostly neglected in research apart from Stack Overflow mining for which the work of Bacchelli et al.~\cite{Bacchelli:2012:HSO:2666719.2666725} is an example. They integrated information from Stack Overflow into the the Eclipse IDE enabling the developer to seamlessly interact with the platform as well as add comments and links to the source code.

\subsubsection{Summary}

All this research aims at only one aspect of the salutogenesis model. In our opinion, this is not enough, because only the combination of all the aspects can lead to an ease of handling the findings of static analysis. Our proposal is not as detailed in each aspect as the  work described above. We aim to seamlessly combine all three aspects, however, to reach a broad spectrum of users by adapting to the personal analysis needs of each user. Also, we believe synergies between the aspects can be utilised to generate new knowledge about the findings.

\section{Transfer of the Model to Static Analysis}
To work successfully with a static analysis tool requires very similar abilities as those described in the salutogenesis model. When triaging the findings of a tool run, the user also has to understand, assess or find meaning in the results provided. It is easy to find examples where these steps are not reached: The description of a finding can be misleading, it is hard to tell how much time it will consume to fix a finding or the prediction of what influence a specific finding will have is hard. Thus, the salutogenesis model can help us here. 
To transfer the model from the medical view to the world of findings of static analysis tools, we have to define what corresponds to the aspects comprehensibility, manageability and meaningfulness.

\textbf{Manageability} describes the feeling towards the resources available to a person to meet the needs raised by external stimuli.
The resources we use in the case of static analysis tools and their findings are time, attention\footnote{Attention can be treacherous: For example, Simons and Chabris~\cite{simons1999gorillas} describe how obvious information can get completely lost, if we concentrate  too much on something else.} and decision making capacity. For decision making capacity, Vohs et al.~\cite{vohs2014making} show in six experiments that choosing drains a psychological resource that is also needed for self-control and taking initiative. So, making many decisions makes us passive and less self-controlled.  Also the probability of making mistakes increases with the cognitive exhaustion of a person as shown by Boksem, Meijman and Lorist~\cite{Boksem2005107}. They had people perform a visual task for 3 hours without a break. Using an EEG measurement, they were able to show that mental fatigue results in a reduction in goal-directed attention.

The amount of findings and options also influences this resource as more choices do not lead to more joy after a certain threshold is reached. In fact, it can have a negative impact going to the extent that the person confronted with these choices refuses to choose at all\cite{huffman1998variety}. 

To save resources, we can reduce the information the user is confronted with and, so, effectively decrease the number of choices and the attention needed. The challenge is to achieve this without reducing comprehensibility and meaningfulness. 
Manageability could also be increased if we could provide an estimated time needed to fix a certain finding of the static analysis tool, as this will enable users to plan their resource usage better.

\textbf{Comprehensibility} describes the extent to which external stimuli can be processed by a person meaning that the information is ordered, consistent, structured and clear.
This principle applied to static analysis aims at the presentation of the results of an analysis as well as the description of the findings and possible solution examples. Solutions from other developers or a community would also increase comprehensibility, as the user can lean on similar solutions when searching for a solution to his or her problem.

\textbf{Meaningfulness} describes a person's understanding of the worth of investing energy in something and to see a problem rather as a challenge than a burden.
We can apply this principle to the tools for static analysis and their findings by adding context to the findings, such as comments or information about human-detected false positives, but also by making the tool's ranking of the findings more understandable or giving the findings a ranking at all. The importance and seriousness of each finding of static analysis tools varies for each user. Some might just accept the rather arbitrary classification of the tool, but in our experience\footnote{e.g. made in the experiment described in\cite{ostberg2013novel} } the more advanced user tends to ignore this classification. This classification should be editable to represent the users state of mind and make the salutogenesis model more fitting the user.

The context added can be extra data provided by other tools. Quality metrics can provide part of this context. It would create additional meaning if we were able to show how a metric develops while removing findings of the static analysis tool. For example, it would be interesting to know, if and how coupling and cohesion are affected by certain findings or if some findings are more likely to be connected to complexity then others. 

Researchers have worked on quality models for several decades to better capture what software quality is. This resulted in a large 
number of available quality models \cite{2009_klaes_QM_landscape}. Quality models could, in principle, constitute a valuable source 
for providing meaning to static analysis findings in the form of the effect on quality. In many existing standard quality models, however, there is a gap between the high-level 
quality attributes and concrete quality measurement \cite{2009_wagners_quality_models_practice}. Modern quality 
models bridge that gap. For example, \emph{Squale} \cite{MordalManet.2009} and \emph{Quamoco}~\cite{wagner:ist15} are operationalised 
quality models that provide a clear rationale for the impact of static analysis findings on quality attributes. Therefore, we can extract that
information for the developer to create meaning.

\section{Integration of the Model into FindBugs}
In the following, we will show how the transfer discussed above can be implemented in a real tool. We choose FindBugs, because it is well known and includes complex findings which are hard to explain and present. We are working on an implementation of this model for FindBugs with the working title ''HaST" (\textbf{H}istory \textbf{a}nd \textbf{S}uggestion \textbf{T}ool).

\subsection{Comprehensibility}
The aim of comprehensibility is to ease the processing of external stimuli. We interpret it here as everything that helps to understand the current finding. 

\begin{figure}
	\centering
	\includegraphics[width=0.45\textwidth]{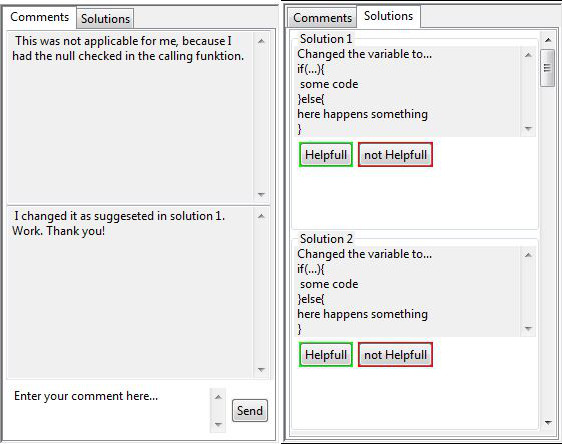}
	\caption{Comments  and solutions as tabs in FindBugs}
	\label{comments}
\end{figure}

Comments of colleagues or other developers can help to understand the core of the finding presented by FindBugs.
Building a repository of comments to the findings of FindBugs\footnote{http://findbugs.sourceforge.net/bugDescriptions.html} can help the user understand what the addressed problem is, especially with intricate findings like multi-threading issues. These comments could also include standard solution approaches or solutions for fixing this finding by other people in other contexts.
The tab for the comments (Fig.~\ref{comments} on the left) includes a text field so the user can add his or her own comment. The tab for the solutions (Fig.~\ref{comments} on the right) includes two buttons for rating the example. This way, we can ensure a minimal quality of the examples. Moreover, the example with the most positive votes should be listed first and examples with only negative votes can be removed after a certain time. 
The comments and solutions will be stored at a central server which spreads the information to the connected clients. The information could be entered anonymously or connected to a user ID. One could extend this by exploiting existing work on Stack Overflow
mining \cite{Bacchelli:2012:HSO:2666719.2666725} to get related comments and rankings.

The rearrangement of the findings also adds to comprehensibility, but we decided to put this into the manageability part of the model. Our focus here is on the reduction of information, and we see the increase in comprehensibility as a pleasant side effect and
benefit of using the model.

\subsection{Meaningfulness}
By increasing the meaningfulness, we show why it is important to work on a finding. We propose three functions, partly based on observations and personal experience, which we will present in the following.

One way to increase the meaningfulness of distinct findings is by connecting them with metrics. To provide meaning, the
metrics themselves should be meaningful for a developer. We would expect that established metrics such as coupling and
cohesion are good candidates. The metrics are gathered by HaST over time with each new FindBugs analysis, and with a significant amount of data it should be possible to predict the impact of the removal of a finding, but also which findings you should work on to improve a certain metric. 

FindBugs already defines severity and confidence of the finding, but this is not visible enough. Here we propose a clearer colour scheme. The colour for the severity rank of a finding is shown prominently in the tree view. The level of confidence is represented by the transparency of the colour, from opaque for high confidence to just shaded for low confidence. False positives will have a colour not to be confused with severity or confidence. We need to find out by an experiment  if we want the false positives to have a colour that stands out to remind the user, that he or she chose to mark this as a false positive and might reconsider, or if we want the false positives to nearly vanish by colouring them grey for example.

\subsection{Manageability}
In our model the increase in manageability is mainly achieved by the reduction of the information that is needed to be processed in a particular moment. We use information levels because we expect them to be convenient and easy to understand for the users.
At the moment we have designed 5 levels because we have found 5 reasonable ways to reduce the information. With increasing level number we cut away more and more  information and focus on the most pressing issues. 
\emph{Level 0} is the original representation of FindBugs.

\begin{figure}
	\centering
	\includegraphics[width=0.3\textwidth]{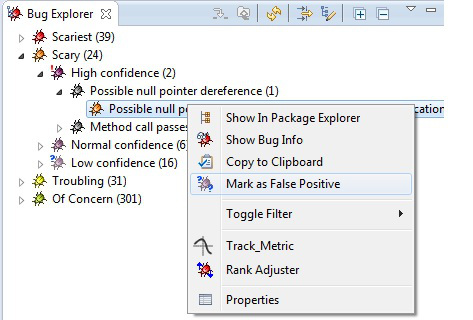}
	\caption{Context menu for removing false positives}
	\label{false_pos_remove}
\end{figure}
On \emph{level 1} we remove findings marked as false positives from the view. Marking false positives is possible at this level by an added context menu (see Fig.~\ref{false_pos_remove}). The finding marked as a false positive is now half-transparent. The marking of the false positives needs to be reliable to be of use. We extended the mechanism already used by FindBugs to store its findings per project in Eclipse.

\begin{figure*}
	\centering
	\includegraphics[width=\textwidth]{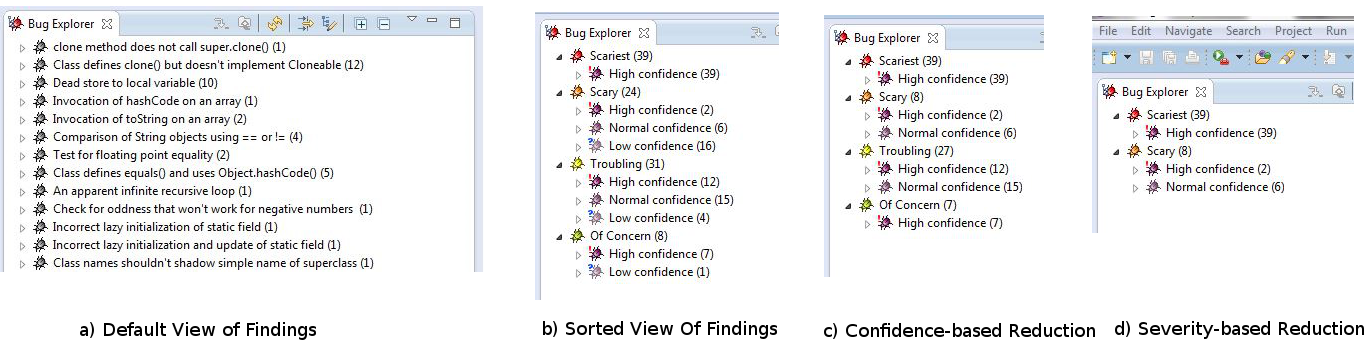}
	\caption{Different stages of sorting of FindBugs findings}
	\label{all_sort}
\end{figure*}

Activation of \emph{level 2} will rearrange the list of findings (see Fig.~\ref{all_sort}a for an example of the default sorting) by severity and confidence (Fig.~\ref{all_sort}b), so that the findings with the highest rank and the highest confidence are now sorted top-most and the rest is listed in descending order (highest rank/mediocre confidence, highest rank/low confidence and mediocre rank/highest confidence).

On \emph{level 3}, we reduce the amount of findings based on the confidence of the finding (e.g. starting from Fig.~\ref{all_sort}b to Fig.~\ref{all_sort}c). A combo box is available to select the level of confidence.

On \emph{level 4} we reduce the amount of findings based on the severity of the finding (e.g. starting from Fig.~\ref{all_sort}c to Fig.~\ref{all_sort}d).  A slider is available for adjusting the confidence level.

The functionality used at levels 2, 3 and 4 is already integrated in FindBugs, but not as accessible -- the controls for this are originally buried deep in the preferences -- and integrated in a self-contained model as in HaST which combines many different information sources.

At \emph{level 5}, we reduce the amount of results shown to a maximal amount of 8 findings from the pool of findings defined by levels 4 and 3. If there are not enough findings in this pool to fill the 8 slots, we loosen the restrictions first to level 4 and then to level 3 until we can deliver 8 findings or there are no more findings left, even after the relaxation of all levels.
The best amount of findings shown at level 5 has to be evaluated by experiments as well as whether the 5 levels are enough.
For example, we could imagine a level 6 where only one finding is picked at random from the same pool of findings as for level~5. 

The severity of single findings should be editable for the users, as they may perceive the severity of findings differently than the tool, because they know their code in depth. Some developers might acknowledge the presence of a finding but with their knowledge of the code, they do not think it is that severe. So they have the possibility to rank it down and deal with it later. This will  help manage the workflow.

To predict the time needed to fix a certain class of findings, we will give the user the opportunity to enter a fix time for each resolved finding after an analysis. This has to be done as non-intrusive as possible. For example, we can trigger a reanalysis on the file that has been worked on at every saving of the file and extract the time of now solved issues. This information will be stored anonymously on the central server so it can be combined with the information from other sources and developers, but cannot be used to assess the performance of single individuals. As the database grows, the prediction will be getting precise enough for estimates within an acceptable uncertainty range ($\pm20$ minutes).

\begin{figure}
	\centering
	\includegraphics[width=0.38\textwidth]{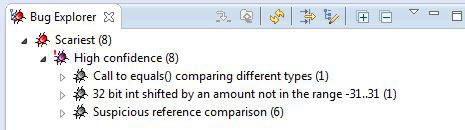}
	\caption{Reduction of findings to a minimal amount}
	\label{magic}
\end{figure}

\subsection{Application of the Model Features}
The user should decide how much control he or she wants to grant the model. Thus, we see the need for user feedback. In the following,
we describe how this feedback can drive the use of the model's features.

\subsubsection{Full Support Start}
The idea is to start with all the features at maximum. The user then loosens the restraints of the model on the findings, until he or she reaches a convenient level.

\subsubsection{No Support Start}
The ``No Support Start" starts with the plain old FindBugs and no features of the model activated. A quick tutorial shows the user how to start and he or she will then enable more and more features until a convenient level is reached.

\subsubsection{Learning as you go}
The basic idea here is that the model will apply a feature and then detect if this application has changed the user's interaction for the better.

 We recommend levels, were applicable, and switches to control every aspect of the model's influence on the original presentation, but also delivering suggested pre-configured steps for convenient use. 

\section{Evaluation}
To evaluate the model we plan an experiment. The participants will get a task to work on a given code base which contains several issues reportable by FindBugs. One group of participants will get FindBugs with HaST; a control group will only have access to the standard FindBugs. Then we can compare the amount of issues fixed by both groups in a fixed amount of time. The main hypothesis is that the group supported by our new model will fix more issues in the same amount of time than the group without the support by the model. 

\section{Conclusion and Future Work}
It is possible to transfer the medical-psychological model of \emph{salutogenesis} into a model for the triage of results of static analysis. In contrast to other approaches it does not treat the three aspects \emph{comprehensibility}, \emph{meaningfulness} and \emph{manageability} separately but in combination. Isolated consideration of these aspects can have disadvantages. E.g., if only meaningfulness is considered, the meaningful findings can get lost in the effort or if only meaningfulness is considered, the result might be unmanageable.  
The aspects are closely connected and each person has a different need for each aspect. We give the user full control over these three aspects. By this, we hope to achieve a positive, stress-free, individual working environment which should lead to higher quality work, faster. As a side effect we will be able to better understand the connections between the findings of FindBugs and other meta information such as metrics.

Future work will include empirical studies on the different aspects of the model. The results of this should help us decide how the different aspects should be integrated and might bring  new ideas and extensions to the model.






%
\bibliographystyle{abbrv}

\end{document}